\documentclass[12pt]{iopart}

\usepackage{amssymb,graphicx,algorithm,algpseudocode,color}

\newcommand{\R}{{\mathord{\mathbb R}}}
\newcommand{\N}{{\mathord{\mathbb N}}}

\newcommand{\dt}{{\rm d}t}
\newcommand{\ds}{{\rm d}s}

\newcommand{\dW}{{\rm d}W}
\newcommand{\vdot}{\dot{V}}
\newcommand{\Vtilde}{{\tilde V}}

\newcommand{\Chat}{\widehat{C}}
\newcommand{\thsp}{\hspace*{0.1ex}}
\newcommand{\tobe}{\mathop{=}\limits^{!}}
\newcommand{\AAA}{\mathbf{A}}
\newcommand{\FF}{\mathbf{F}}
\newcommand{\kernel}{\mathbf{K}}
\newcommand{\friction}{\mathbf{K}_0}
\newcommand{\ud}{\textrm{d}}

\renewenvironment{matrix}{%
  \matrix@check\matrix\env@matrix
}{%
  \endarray \hskip -\arraycolsep
}

\makeatletter
\newcount\c@MaxMatrixCols \c@MaxMatrixCols=10
\newenvironment{cmatrixc}{%
  \hskip -\arraycolsep\array{*\c@MaxMatrixCols c}%
}{%
  \endarray \hskip -\arraycolsep
}
\makeatother

\newenvironment{cmatrix}{\left[\cmatrixc}{\endmatrix\right]}

\begin{document}

\title[Model reduction for generalized Langevin equations]{Model reduction 
techniques for the computation of extended Markov parameterizations for 
generalized Langevin equations}

\author{N Bockius$^1$, J Shea$^2$, G Jung$^3$, F Schmid$^2$, M Hanke$^1$}
\address{$^1$ Institut f\"ur Mathematik, Johannes
    Gutenberg-Universit\"at Mainz, 55099 Mainz, Germany}
\address{$^2$ Institut f\"ur Physik, Johannes
    Gutenberg-Universit\"at Mainz, 55099 Mainz, Germany}
\address{$^3$ Institut f\"ur Theoretische Physik, Universit\"at
Innsbruck, Technikerstraße 21A, A-6020 Innsbruck, Austria}
\ead{hanke@math.uni-mainz.de}

\begin{abstract}
The generalized Langevin equation is a model for the motion of coarse-grained
particles where dissipative forces are represented by a memory term. 
The numerical realization of such a model requires 
the implementation of a stochastic delay-differential equation and the
estimation of a corresponding memory kernel. Here we develop
a new approach for computing a data-driven Markov model for the motion of the
particles, given equidistant samples of their velocity autocorrelation
function. Our method bypasses the determination of the underlying
memory kernel by representing it via 
up to about twenty auxiliary variables.
The algorithm is based on a sophisticated variant of the Prony 
method for exponential interpolation and employs the Positive Real Lemma from
model reduction theory to extract the associated Markov model. 
We demonstrate the potential of this approach for the test case of anomalous
diffusion, where data are given analytically, and then apply our method to
velocity autocorrelation data of molecular dynamics simulations of a colloid
in a Lennard-Jones fluid. 
In both cases, the VACF and the memory kernel can be reproduced
very accurately. Moreover, we show that the algorithm can also handle 
input data with large statistical noise. We anticipate that it will be
a very useful tool in future studies that involve dynamic coarse-graining 
of complex soft matter systems.
\end{abstract}


\submitto{\JPCM}

\maketitle

\sloppy

\section{Introduction}
\label{Sec:Intro}

Generalized Langevin Equations (GLE)s, i.e., extensions of Langevin
equations with memory, have fascinated scientists for many decades,
ever since they were first introduced by Mori in the 60s \cite{Mori65} 
based on concepts of Zwanzig~\cite{Zwanzig61}. Their purpose
is to describe in very general terms the irreversible dynamics of 
collective (coarse-grained) observables in many-particle systems 
without having to assume complete separation of time scales. 
Consider a multiparticle system at thermal equilibrium, and
assume we are interested in the dynamical evolution of a given set
$\AAA(t)$ of coarse-grained variables. For such cases, starting from a
microscopic Hamiltonian description and using a projection operator
formalism, Mori and Zwanzig derived a dynamical equation of the
following form for $\AAA(t)$ \cite{Zwanzig_book}
\begin{equation}
 \dot\AAA(t) = \FF^C(\AAA(t))
         - \int_0^t \ds\, \kernel(t-s) \AAA(s) + \FF^R(t).
          \label{eq:MZ-GLE}
\end{equation}

Here $\FF^C(\AAA(t)) = \textrm{i} \Omega \AAA $ describes the collective
oscillations of $\AAA(t)$, $\kernel(t)$ a memory kernel matrix, and
$\FF^R(t)$ incorporates nonlinear terms and interactions with the
remaining microscopic degrees of freedom and is interpreted as a
stochastic process. It is connected to the memory kernel via a
fluctuation-dissipation relation 
\begin{equation}
 \langle \FF^R(t) \FF^R(t') \rangle = 
   \kernel(t-t') \langle \AAA \AAA \rangle, 
          \label{eq:MZ-FDT}
\end{equation}
where $\FF^R(t) \FF^R(t')$ and  $\AAA \AAA$ denote tensor products 
and $\langle \cdot \rangle$ configurational averages. In the limit where
the memory kernel decays almost instantaneously, Eq.\ (\ref{eq:MZ-GLE}) 
can be replaced by a standard Markovian Langevin equation with 
friction $\friction = \int_0^\infty \ud s \: \kernel(s)$ and uncorrelated 
noise $\FF^R(t)$ satisfying $\langle \FF^R(t) \FF^R(t') \rangle = 
2 \friction \delta(t-t') \langle \AAA \AAA \rangle$.

Eq.~(\ref{eq:MZ-GLE}) can be generalized to stationary and even
non-stationary nonequilibrium systems \cite{Grabert_book,MVS17}.
Using modified projector methods, it is possible to design GLEs such
that $\FF^C$ includes all (linear and nonlinear) reversible
interactions \cite{KH07}, and the memory and noise terms subsume the
remaining dissipative contributions to the dynamical equations. In
that case, the latter can also be seen as a frequency dependent
thermostat \cite{Zwanzig_book}. Independent of the original dynamic
coarse-graining context, such GLE thermostats have been utilized to
devise enhanced
sampling approaches in molecular dynamics simulations
\cite{CBP09} and even as means to mimic the effect of nuclear
quantum fluctuations \cite{CBP09b}. From a more fundamental point
of view, GLEs are popular theoretical frameworks to describe
intriguing physical phenomena such as anomalous
diffusion \cite{MK00,HF13,Goychuk12} or the glass 
transition \cite{Goetze_book}.

In numerical simulations of GLEs, one faces two main challenges:
First, the efficient evaluation of the history dependent memory term,
and second, the generation of suitably correlated random numbers for
the stochastic term. If the memory kernel has a finite range, a
straightforward direct integration of the GLE is
possible \cite{EB80,SH91,CLL14,LLDK17,JHS17,JHS19}. However, depending on
the memory range, the integration can be time consuming, and the
necessity to impose a sharp cutoff may cause problems. Therefore,
already early on, alternative approaches have been explored where the
GLE is approximated by an extended Markovian system, either using autoregressive techniques \cite{Mori65b,CR80,FP79,MG83} or by introducing
additional, auxiliary degrees of freedom \cite{CBP10,BB13,SLK14,MLL16a}. 
In both cases, the memory kernel is effectively approximated by a 
sum of exponentials.
Integrating GLEs {\em via} extended Markovian schemes has
several advantages over direct integration.  First, the hard cutoff of
the memory kernel is replaced by a less severe exponential cutoff.
Second, the integration is often faster \cite{LLDK17}. Third, the
integration scheme can be implemented in a straightforward manner
using established algorithms for Langevin equations.

To simplify the discussion, in the following, we will focus on
GLEs that describe the Brownian motion of selected tagged particles
(e.g., colloids or polymers) in a bath of surrounding particles. The
coarse-grain quantity of interest is the 
velocity $V(t)$ of the center
of mass of the tagged particle. Our considerations and methods can
easily be generalized to other GLEs as well. The derivation of extended
Markov systems for approximating the solution of such GLEs has been
treated in several papers. The standard work flow consists of three
steps.  In a first step an approximation of the memory kernel is
determined.  In some cases, this is done directly by using the
Mori-Zwanzig formalism analytically \cite{CLL14,MLL16}, or by
analyzing trajectories in microscopic simulations with concepts from
the Mori-Zwanzig theory \cite{CVR14}. More often, one uses known velocity
and force correlation functions from all-atom simulations
\cite{JHS17,JHS19,LBL16,LLDK17,YLKK17} as input data and solves a first kind
or second kind Volterra integral equation.
Care has to be taken in this latter case, because first kind integral
equations are notoriously ill-conditioned \cite{Brunner17,KDKSSN19,SKTL10}.  
In a second step the computed memory kernel is approximated by an
exponential series, known as a Prony series; this can be achieved by
using rational approximations of the Laplace transform of the memory
kernel, or by some general nonlinear fitting scheme
\cite{BB13,FYEF09,LBL16,MLL16}; both methods, however, are only
feasible for few terms of this exponential series. Finally, the
coefficients of the extended Langevin model can be extracted from the
coefficients of the exponential approximation
\cite{CBP10,FYEF09,Pavliotis14}.

In case the starting point is the velocity autocorrelation function (VACF), 
determining the memory kernel prior to expanding it
seems an unecessary detour. It should be more efficient and less
error-prone to extract and optimize the parameters of the integrator
directly from the VACF data.  Wang \emph{et al.} have taken this
direct route and determined the parameters of a Prony series with up
to eight terms by training a deep neural network in two steps with VACF
data \cite{WMP20}. Using the example of dilute and dense star polymer
solutions, they showed that their GLE integrator is able to reproduce 
the early and late time characteristics of the VACF over several
orders of magnitude in time. This example demonstrates the potential
of direct optimization. However, their machine learning based method 
requires a complicated training procedure involving a series of
GLE simulations, similar to other iterative memory reconstruction 
methods \cite{JHS17}. Therefore, it gives little insight into the
mathematical structure of the problem, and in particular, the 
impact of noise in the input data remains unclear.

In the present paper, we
propose an analytical method to directly determine the Langevin model
from a finite number of samples of the VACF.
We avoid taking the detour of computing an intermediate memory kernel
and thus circumvent a significant cause of numerical errors.  Our
approach adapts techniques which have originally been developed for
the purpose of model reduction in deterministic dynamical
systems~\cite{FF95,Freund03}. Some of these have already been used in
the context of stochastic dynamical systems~\cite{BR15,FF03}, but here
we follow a different route.

The paper is organized as follows. In Section~\ref{Sec:Methods}, we describe
the background and major ideas behind our model reduction algorithm;
more technical details have been shifted into three appendices at the 
end of this text.
Then, in Section~\ref{Sec:Results},
we discuss numerical results for three different case studies:
one setting uses analytic data (``subdiffusion''), a second example treats
molecular dynamics (MD) data taken from our earlier paper \cite{JHS17},
and the final case study considers a sequence of MD data sets with 
different amounts of statistical noise.
A final discussion of our results in Section~\ref{Sec:Conclusion}
concludes this work.

\section{Methods}
\label{Sec:Methods}
In this paper we restrict ourselves to a one-dimensional system,
and postpone the nontrivial technical generalization to higher
dimensional problems to future work. 
We thus consider specifically the GLE
\begin{equation}
\label{GLE}
   \ m\vdot(t) \,=\, -\!\int_{-\infty}^t\ds\,\gamma(t-s)V(s) \,+\, F^R(t)
\end{equation}
for the scalar velocity $V$ of a macroparticle with mass $m>0$,
where the memory kernel $\gamma$ is taken to be continuous and integrable. 
The lower bound of the integral in Eq.~(\ref{GLE}) is chosen $-\infty$
to indicate that we are considering the stationary limit of a process
where the origin of time does not matter.
In order to fulfill the fluctuation-dissipation theorem the random force 
$F^R$ 
is assumed to be a stationary Gaussian process with zero mean, satisfying
\[
   \ \langle F^R(t)F^R(t')\rangle \,=\, \frac{1}{\beta}\,\gamma(t-t')\,, 
\]
where $\beta$ is the inverse temperature~\cite{Pottier10}. 
The corresponding solution $V$
is a stationary Gaussian process with zero mean and autocorrelation function
\begin{eqnarray}
\label{VACF}
   \ C_V(t-t') \,=\, \langle V(t)V(t')\rangle\,.
\end{eqnarray}

We assume that samples 
\begin{equation}
\label{yk}
   \ y_\nu \,=\, \beta m\, C_V(t_\nu)
\end{equation}
of this autocorrelation function are given on an equidistant grid 
\begin{eqnarray}
\label{grid}
   \ \triangle \,=\, \{t_\nu \,=\, \nu\tau\,:\, \nu=0,\dots,2n-1\}
\end{eqnarray}
for some $n\geq 2$ and mesh size $\tau>0$. Due to the particular specification
(\ref{yk}) these data are normalized to satisfy \mbox{$y_0=1$} up to 
statistical errors. In the initialization step of our algorithm
we therefore rescale the data to exactly match this physical constraint.

Our goal is the derivation of a Langevin equation
\begin{eqnarray}
\label{Langevin}
   \ {\rm d}\!\begin{cmatrix} \Vtilde \\ Z \end{cmatrix}
   \,=\, k 
         \begin{cmatrix}
            \phantom{-}0\phantom{_1} & \!\thsp e_1^T\\ \!-e_1 & \!T_0 
         \end{cmatrix}
         \begin{cmatrix} \Vtilde \\ Z \end{cmatrix}\dt
       + \begin{cmatrix} \,0\,\\ g \end{cmatrix} \dW
\end{eqnarray}
for an approximate velocity $\Vtilde$ and a set of auxiliary variables 
$Z\in\R^N$, with $e_1$ being the first canonical basis vector in $\R^N$ and
$W$ a one-dimensional Brownian motion.
We are interested in the stationary solution of (\ref{Langevin}), and
we want to achieve $\Vtilde \approx V$ by choosing the coefficients 
$k\in\R$, $g\in\R^N$, and the tridiagonal matrix $T_0\in\R^{N\times N}$ 
in such a way that the autocorrelation function $C_\Vtilde$ 
of $\Vtilde$ is given by a finite Prony series
\begin{eqnarray}
\label{Prony}
   \ C_\Vtilde(t) \,=\, \frac{1}{\beta m}\sum_{j=1}^{N+1} w_j e^{\lambda_jt}, \quad
   t\geq 0\,,
\end{eqnarray}
and interpolates the given data at the $2n$ grid points of $\triangle$, i.e.,
\begin{equation}
\label{interpol}
   \ \beta m\, C_\Vtilde(t_\nu) \,\approx\, y_\nu\,, \quad \nu=0,\dots,2n-1\,.
\end{equation}

We mention that if the memory kernel is of interest for diagnostic purposes
then it can readily be computed as
\begin{equation}
\label{memory}
   \ \gamma(t) \,=\, k^2 e_1^Te^{t\thsp kT_0}e_1\,, \quad t\geq 0\,.
\end{equation}

\subsection{Prony's method}
\label{Subsec:Prony}
In the first step of our approach we determine the solution
\begin{equation}
\label{f}
   \ f(t) \,=\, \sum_{j=1}^{n} w_j e^{\lambda_jt}
\end{equation}
of the exponential interpolation problem
\begin{equation}
\label{expinterpol}
   \ f(t_\nu) \,=\, y_\nu\,, \quad \nu=0,\dots,2n-1\,.
\end{equation}
For this we use a variant of Prony's method, which we borrow
from \cite{ADM91}. This algorithm computes 
the 
Lanczos polynomials associated with the moment functional
defined by the given data (see \ref{App:Algorithm} for more details)
and uses the corresponding recursion coefficients
to set up a (real) tridiagonal Jacobi matrix $J\in\R^{n\times n}$.
The key observation behind the algorithm is that this matrix solves the
moment problem
\begin{equation}
\label{Ammar}
   \ e_1^T J^\nu e_1 = y_\nu\,, \quad \nu=0,\dots,2n-1\,,
\end{equation}
where now $e_1 = [1,0,\dots,0]^T$ is the first canonical basis vector in $\R^n$.

We should mention that the interpolation problem (\ref{f}), (\ref{expinterpol})
need not have a solution, and that this recursive algorithm can break down,
but we never encountered this problem in our experiments.
We also assume in the sequel that $J$ has $n$ distinct
eigenvalues $\mu_j$ to simplify the presentation. Then we can factorize
\begin{eqnarray}
\label{Jn}
   \ J \,=\, XDX^{-1}\,,
\end{eqnarray}
where $D$ is a diagonal matrix with the eigen\-values $\mu_j$ on its diagonal; 
the columns $x_j$ of 
\begin{eqnarray}
\label{X}
   \ X \,=\, [x_1,\dots,x_n]
\end{eqnarray}
are the associated eigenvectors. Given this factorization we then let
\begin{equation}
\label{A}
   \ A \,=\, \frac{1}{\tau}\log J \,=\, X \Lambda X^{-1} \,,
\end{equation}
where $\Lambda\in\R^{n\times n}$ is the diagonal matrix with the eigenvalues 
\begin{equation}
\label{lambda}
   \ \lambda_j \,=\, \frac{1}{\tau}\log\mu_j
\end{equation}
of $A$ on its diagonal. It follows that $e^{\nu\tau A} = J^\nu$ for $\nu\in\N_0$, 
and hence,
\begin{equation}
\label{expinterpolsol}
   \ f(t) \,=\, e_1^Te^{tA}e_1
\end{equation}
is the desired solution of the interpolation problem (\ref{expinterpol}).

When the memory kernel is integrable and continuous 
then the autocorrelation function $C_V$ is differentiable with
\[
   \ \dot{C}_V(0) \,=\, 0\,.
\]
We therefore also like to impose the condition $\dot{f}(0)=0$ for the
Prony series (\ref{expinterpolsol}), which gives
\begin{equation}
\label{A11}
   \ 0 \,\tobe\, \dot{f}(0) \,=\, e_1^TAe_1 \,=\, A_{1,1}\,.
\end{equation}
In general, however, the matrix $A$ generated in (\ref{A})
does not satisfy this constraint. Our remedy for this shortcoming is 
that we blame this on measurement errors in $y_1=\beta m\,C_V(\tau)$, 
i.e., the first nontrivial data point,
and we seek to perturb $y_1$ in such a way that (\ref{A11}) is satisfied.
We apply a Newton iteration for this purpose, which is described in
\ref{App:Newton}.

We emphasize that the problem of fitting exponentials to given data is known
to be highly ill-conditioned \cite{Varah85}. In our context this is 
reflected by the presence of 
spurious exponentials in (\ref{f}) 
with exponents $\lambda_j$ in the closed right-half plane,
corresponding to eigenvalues $\mu_j$ of $J$ in the exterior of the unit disk 
or on the unit circle.
The corresponding exponentials are unphysical because they do not converge to zero.
Fortunately, though, if $\tau$ has been properly chosen and the data noise is
not too big (see Section~\ref{Sec:Results} below), 
then the respective terms in (\ref{f}) come with weights, which are by several 
orders of magnitude smaller than those of the relevant terms. 
We therefore delete the spurious terms from the series representation of $f$
defined in (\ref{expinterpolsol}) by eliminating the corresponding eigenvalues.
This modification is described in more 
detail in \ref{App:SpectralModifications}. 

Another issue that is worth mentioning concerns negative eigenvalues $\mu_j$
(within the unit disk) of the Jacobi matrix -- in particular, because this
detail is not addressed at all in \cite{ADM91}. The complex logarithm which is
employed in (\ref{lambda}) is defined in the complex plane except for a cut
along the nonpositive real axis. However, the presence of eigenvalues 
$\mu_j\in(-1,0)$ makes sense physically; in view of (\ref{Ammar}) those 
correspond to damped oscillations in the data, and hence, to terms of the
form 
\begin{equation}
\label{negative-term}
   \ w_j|\mu_j|^{t/\tau}\!\cos(\pi t/\tau)
\end{equation}
in (\ref{f}).
As the logarithm is multivalued for negative arguments 
we need to replace the corresponding formula (\ref{lambda}) by adding 
\emph{two} eigenvalues
\begin{equation}
\label{lambda-special}
   \ \lambda_{\pm j} \,=\, 
   \frac{1}{\tau}\log|\mu_j| \,\pm\, \frac{\pi}{\tau}\,\rmi
\end{equation}
to the spectrum of $A$ for a correct representation of this term 
in (\ref{expinterpolsol}).
Again, we refer to \ref{App:SpectralModifications} to illustrate
how this can be achieved.

The final matrix $A$ has the form
\begin{eqnarray}
\label{A-blocks}
   \ A \,=\, \begin{cmatrix} \phantom{-}0 & b^T\\ -c & A_0 \end{cmatrix}.
\end{eqnarray}
It is no longer $n\times n$, but typically has a smaller dimension 
$(N+1)\times(N+1)$, because of the elimination of the spurious eigenvalues.

\subsection{The Positive Real Lemma}
\label{Subsec:PRL}
In a second step of our algorithm we determine a vector $\ell\in\R^N$, such
that the auxiliary Langevin equation
\begin{eqnarray}
\label{Langevin-tmp}
   \ {\rm d}\!\begin{cmatrix} \Vtilde \\ Y \end{cmatrix}
   \,=\, \begin{cmatrix} \phantom{-}0 & b^T\\ -c & A_0 \end{cmatrix}
     \begin{cmatrix} \Vtilde \\ Y \end{cmatrix}\dt
     + \begin{cmatrix} \,0\,\\ \ell \end{cmatrix} \dW
\end{eqnarray}
with the system matrix determined in (\ref{A-blocks}) 
has a stationary solution, 
whose autocorrelation function $C_\Vtilde$ is given by (\ref{Prony}).
Before doing so let us make two remarks.

\begin{enumerate}
\item 
Due to our construction the matrix $A$ of (\ref{A-blocks}) has only eigenvalues 
with negative real part. 
Therefore, the Langevin equation (\ref{Langevin-tmp}) has a unique 
stationary solution, and the corresponding  autocorrelation function 
$C_\Vtilde$ is given by
\begin{equation}
\label{exptAX}
   \ C_\Vtilde(t) \,=\, \frac{1}{\beta m}\,e_1^T e^{|t|\!\thsp A}\Sigma\,e_1\,,
\end{equation}
where the covariance matrix $\Sigma$ is the solution of the 
Lyapunov equation
\begin{equation}
\label{Lyapunov}
   \ A\Sigma \,+\, \Sigma A^T \,=\, 
   - \beta m \begin{cmatrix} 0 & 0\, \\ 0 & \ell\ell^T \end{cmatrix},
\end{equation}
cf., e.g., \cite{Pavliotis14}.
Since we want $\beta m\,C_\Vtilde(t)$ to match the Prony series $f(t)$ 
for $t\geq 0$ we conclude from (\ref{exptAX}) and (\ref{expinterpolsol}) 
that we need to impose the constraint
\begin{equation}
\label{Sigma-blocks}
   \ \Sigma\,e_1 \,=\, e_1\,,
\end{equation}
i.e., the covariance matrix has to satisfy
\begin{equation}
\label{constraint}
   \ \Sigma \,=\, \begin{cmatrix} 1 & 0\, \\ 0 & \Sigma_0 \end{cmatrix}
\end{equation}
for some symmetric, positive, semidefinite matrix $\Sigma_0\in\R^{N\times N}$.

Inserting (\ref{A-blocks}) and (\ref{Sigma-blocks}) into (\ref{Lyapunov}),
we see that (\ref{Lyapunov}) is equivalent to the system
\begin{equation}
\label{Lure}
\eqalign{\ A_0\Sigma_0+\Sigma_0A_0^T \,=\, -\beta m\,\ell\ell^T\,,\cr
         \ \Sigma_0b \,=\, c\,,}
\end{equation}
which is known as (singular) Lur'e equations. 

\item 
From the Lur'e equations it is evident that a necessary condition 
for the existence of an approximate Langevin system (\ref{Langevin-tmp})
is that all eigenvalues of $A_0$ must have nonpositive real part.

Another necessary condition stems from the well-known fact that
$C_\Vtilde$ is \emph{per se} a function of
positive type, i.e., its Fourier transform $\Chat_\Vtilde$ is nonnegative. 
Since 
\[
  \ \Chat_\Vtilde(\omega) = \frac{2}{\beta m}
    {\rm Re}\Bigl(\rmi\omega + b^T\!(\rmi\omega I-A_0)^{-1}c\!\Bigr)^{\!-1}
\]
by virtue of (\ref{exptAX}) and (\ref{A-blocks}), 
it follows that this condition is satisfied, if and only if
\begin{equation}
\label{posreal}
   \ {\rm Re}\Bigl(b^T(\rmi\omega I-A_0)^{-1}c\Bigr) \,\geq\, 0
\end{equation}
for all $\omega\geq 0$. 

This is a familiar condition in control theory, where the function
\begin{equation}
\label{transfer}
   \ \kappa(s) \,=\, b^T(sI-A_0)^{-1}c
\end{equation}
is known as the transfer function associated with the 
system~(\ref{Langevin-tmp}). Because of our previous statement this function
is analytic in the open right-half plane, and if it also satisfies 
(\ref{posreal}) then it is called positive real. 
Unfortunately, it is hard to deduce from (\ref{f})
whether the transfer function $\kappa$ is positive real.
\end{enumerate}

We have seen that a necessary condition to find a solution of the 
Lur'e equations (\ref{Lure}) is, that the transfer function $\kappa$ be 
positive real. On the other hand a classical result from 
Anderson~\cite{Anderson67a} states that the Lur'e equations do indeed 
have a solution when $\kappa$ is positive real.
Even better, there is also a constructive algorithm for computing 
$\Sigma_0$ and $\ell$; see \ref{App:Algorithm}.
Therefore the computation of the coefficients of the extended Langevin 
equation (\ref{Langevin-tmp}) is settled in the positive real case. 
When this algorithm fails, on the other hand,
then this means that the transfer function (\ref{transfer}) lacks to be
positive real, and then no extended Langevin model (\ref{Langevin-tmp}) exists 
for this particular set of data. Usually this means that 
the grid spacing $\tau$ or the number $2n$ of grid points is either
too big or too small (see Section~\ref{Sec:Results}).

\subsection{A final Lanczos sweep}
\label{Subsec:Lanczos}
The system matrix $A$ of the auxiliary Langevin equation (\ref{Langevin-tmp})
is a full matrix, in general, and hence, the numerical integration
of (\ref{Langevin-tmp}) amounts to $O(N^2)$ operations per time step. 
We therefore perform a coordinate trans\-formation
\begin{equation}
\label{Trafo}
   \ T \,=\, U^{-1}AU
\end{equation}
such that $T$ is a tridiagonal matrix (which reduces
the work load of the time integration to $O(N)$ operations per time step).

The matrix $U$ and its inverse can be determined with the (nonsymmetric)
Lanczos method \cite{Komzsik03}, which generates two vector sequences
$u_0,\dots,u_N$ and $v_0,\dots,v_N$ in $\R^{N+1}$ with
\begin{equation}
\label{biorth}
   \ v_i^Tu_j \,=\, \delta_{ij}\,, \quad i,j=0,\dots,N\,.
\end{equation}
When choosing $u_0$ and $v_0$ as the first canonical basis vector in $\R^{N+1}$ 
then it follows from the biorthogonality relation (\ref{biorth}) that
\begin{equation}
\label{W0}
   \ U \,=\, [u_0,\dots,u_N]
       \,=\, \begin{cmatrix} 1 & 0\, \\ 0 & U_0 \end{cmatrix}
\end{equation}
for some $U_0\in\R^{N\times N}$, and that
\[
   \ U^{-1} \,=\, \begin{cmatrix} 1 & 0\, \\ 0 & U_0^{-1} \end{cmatrix}
   \,=\, [v_0,\dots,v_N]\,.
\]
With $g=U_0^{-1}\ell$ and new variables $Z=U_0^{-1}Y$
the Langevin equation (\ref{Langevin-tmp}) is transformed into
\[
   \ {\rm d}\!\begin{cmatrix} \Vtilde \\ Z \end{cmatrix}
   \,=\, \begin{cmatrix} \,0 & \!\!b^TU_0\\ -U_0^{-1}c & \!\!T_0'\end{cmatrix}
     \begin{cmatrix} \Vtilde \\ Z \end{cmatrix}\dt
     + \begin{cmatrix} \,0\,\\ g \end{cmatrix}\dW
\]
with a tridiagonal matrix $T_0'=U_0^{-1}A_0U_0\in\R^{N\times N}$.
Since $b^Tc>0$ it can be shown that
\[
   \ U_0^T b \,=\, U_0^{-1}c \,=\, k e_1 
\]
with $e_1$ the first canonical basis vector in $\R^N$ and $k=\pm\sqrt{b^Tc}$
(the sign depending on the particular implementation of the Lanczos method),
and hence, we have obtained the desired system~(\ref{Langevin}) with
$T_0=T_0'/k$.

The fact that $b^Tc$ is a positive number can be seen as follows: 
Eq.\,(\ref{Lure}) implies that $b^Tc$ is nonnegative and can only be zero 
when $c=0$; by virtue of (\ref{A-blocks}), however, 
the latter would imply that the matrix $A$ is singular, 
in contradiction to all eigenvalues of $A$ having negative real part. 

Similar to the first step of our algorithm presented in 
Section~\ref{Subsec:Prony} the two vector sequences are computed recursively, 
and the recursion coefficients define the entries of the tridiagonal 
matrices $T$ and $T_0$, respectively. We mention that the Lanczos method can 
encounter a breakdown in exceptional situations \cite{Komzsik03};
for larger values of $N$ the algorithm can also suffer from a loss of 
biorthogonality in (\ref{biorth}) due to accumulated round-off, and this
can lead to new spurious eigenvalues of $T$ in the right-half plane.
When either of this happens, then one has to employ the auxiliary extended 
Langevin equation (\ref{Langevin-tmp}) instead of (\ref{Langevin}).

\section{Numerical results}
\label{Sec:Results}

\subsection{Subdiffusion}
\label{Subsec:slowdiffusion}
We start with a well-known model problem \cite{Lutz01,VD06,HKR16}
known as subdiffusion, where the memory kernel is 
\begin{equation}
\label{memory-subdiffusion}
  \ \gamma(t) \,=\, \frac{1}{\sqrt{\pi}}\,t^{-1/2}\,, \quad t>0\,,
\end{equation}
and the VACF 
\[
  \ C_V(t) \,=\, E_{3/2}(-|t|^{3/2})\,, \quad t\in\R\,,
\]
is explicitly given in terms of the Mittag-Leffler function 
\[
  \ E_\alpha(t) \,=\, \sum_{n=0}^\infty \frac{t^n}{\Gamma(\alpha n+1)}\,.
\]
This setting uses reduced (dimensionless) units, 
so that \mbox{$m=1$} and \mbox{$\beta=1$}.
The knowledge of exact (clean) data allows for a proof of concept of our 
method,
although the memory kernel~(\ref{memory-subdiffusion}) is not integrable and 
exhibits a singularity at $t=0$, whereas our extended Langevin model is always 
associated with a continuous memory kernel, see~(\ref{memory}).

\begin{figure}
\centerline{\includegraphics[height=5cm]{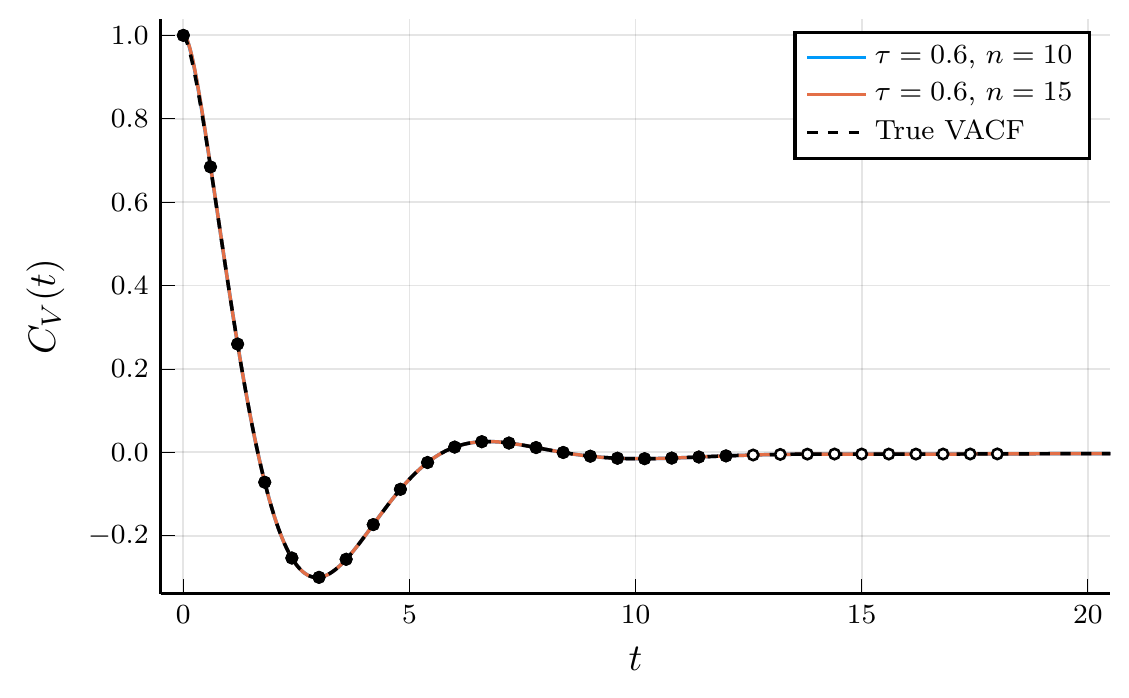}}
\caption{\label{Fig1}The true VACF and two approximate VACFs for the 
subdiffusive model. The black points represent the data samples used in the $[0,12]$ grid to approximate the VACF and the white points represent the additional data samples used on the $[0,18]$ grid.}
\end{figure}

Given the shape of the VACF displayed in 
Figure~\ref{Fig1} we have run our algorithm with data samples
for three different grid spacings, i.e., values of $\tau$; for each 
choice we have compared two grids covering the time interval $[0,12]$ and approximately $[0,18]$, respectively. The corresponding parameter values are
\begin{itemize}
\item $\tau=0.4$ with $n=15$ and $n=22$,
\item $\tau=0.6$ with $n=10$ and $n=15$,
\item $\tau=1.0$ with $n=6$ and $n=9$.
\end{itemize}
In Figure~\ref{Fig1} the target data for $\tau=0.6$ are highlighted 
(the additional data for the longer time interval are marked in white);
besides the exact VACF this figure also shows
the corresponding approximations (\ref{exptAX}) for both associated values
of $n$. In this scale the approximations
exhibit a perfect fit of the true autocorrelation function. The same is true
for all other four aforementioned choices of the two parameters $\tau$ and $n$.

\begin{figure}
\centerline{\includegraphics[width=8.3cm]{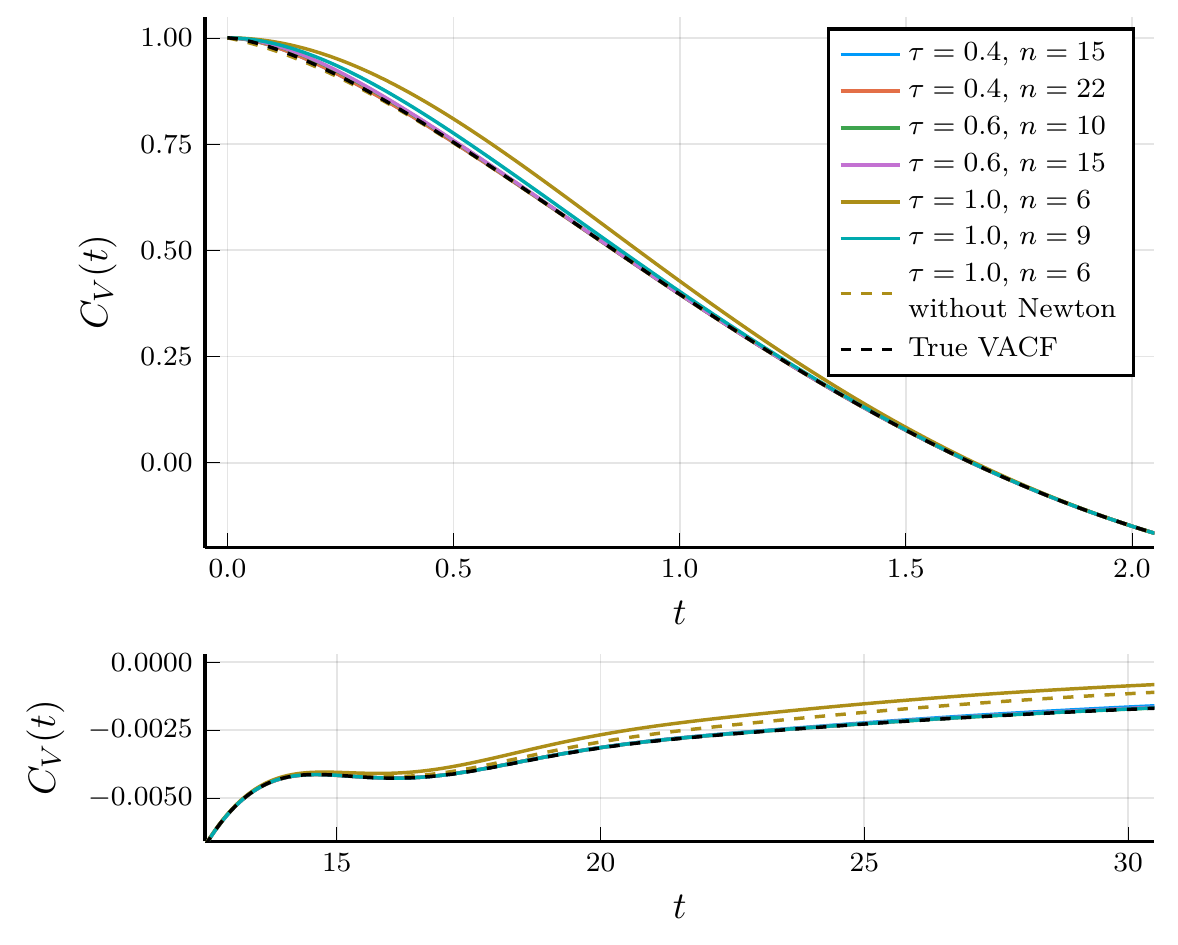}}
\caption{\label{Fig2}Enlargements of the graph of the subdiffusive VACF
and its approximations. The upper panel shows the VACF at short times 
($t=0.0$ to $t=2.0$) and the lower panel shows the same VACF at long times 
($t=10$ to $t=30$).}
\end{figure}

A zoom into the approximate VACFs, however, provides some practical guidelines 
on how to choose $\tau$ and $n$ for a given problem. For example, the
reconstructions corresponding to $\tau=1.0$ exhibit a certain misfit near 
$t=0$ (upper panel in Figure~\ref{Fig2}), and the one for $\tau=1.0$ and $n=6$
also has difficulties in matching the small wriggle near $t=17$ and the 
subsequent tail of the true autocorrelation function 
(bottom panel in Figure~\ref{Fig2}); all curves that are not visible
in these plots lie exactly underneath the true VACF. 
Such misfits show that the grid spacing $\tau=1.0$ 
is too large to cope with finer details and
the infinite curvature at $t=0$ of the true autocorrelation function. 

As a matter of fact, our Prony series approximation for $\tau=1.0$ and $n=6$ 
fails to correspond to a positive real transfer function and does not lead 
to a valid Langevin system. 
For these parameters a modified system (\ref{Langevin}) with nonzeros in the 
(1,1)-entry of the system matrix and the top entry of the noise vector 
can be constructed by dropping constraint (\ref{A11}) (thereby foregoing the Newton iteration),
resulting in $\dot{f}(0)=-0.204$.
The corresponding approximate VACF is included as dashed line in 
Figure~\ref{Fig2}, but is hardly visible in the upper panel, as it is almost
enteirely hidden underneath the true VACF; for larger $t$, though, its fit is not much 
better than the original one according to the lower panel.

\begin{figure}
\centerline{\includegraphics[height=5cm]{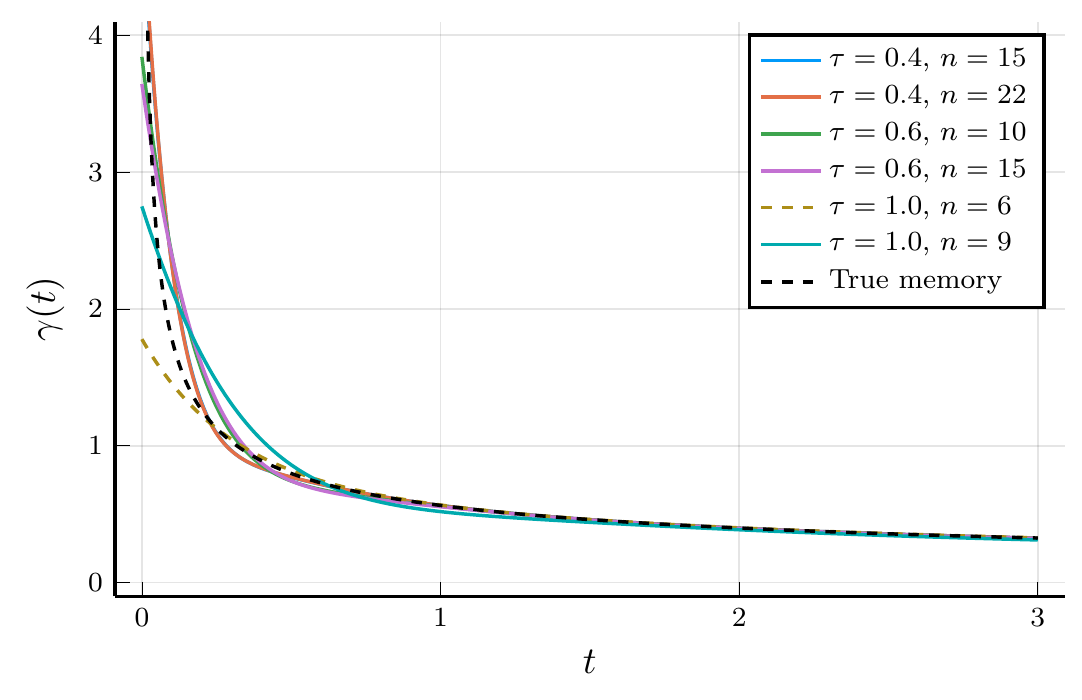}}
\caption{\label{Fig3}Approximations of the subdiffusive memory kernel.}
\end{figure}

Figure~\ref{Fig3} presents the reconstructed 
memory kernels (\ref{memory}) for the various
combinations of $\tau$ and $n$; again, the dashed line corresponds to the
case $\tau=1.0$ and $n=6$ without the constraint (\ref{A11}), where the
memory kernel comes with an additional delta distribution at $t=0$.
It can be seen that the quality of the approximation increases with reduced
grid spacing and with increasing number of grid points.

\begin{table}[h]
\footnotesize
\centerline{\begin{tabular}{@{}c|cccccc}
\br
$\tau$ & \ 1.0 & 1.0 & 0.6 & 0.6 & 0.4 & 0.4 \\
$n$ & \ 6 & 9 & 10 & 15 & 15 & 22 \\
\mr
$N$ & \ {\it 5} & 8 &\phantom{1}9 & 11 & 10 & 11 \\
\br
\end{tabular}}
\caption{\label{Tab1}Number $N$ of auxiliary variables $Z$ in (\ref{Langevin}) 
for different choices of $\tau$ and $n$. The italicized entry $N=5$ in the
first column corresponds to the unconstrained VACF approximation.}
\end{table}

It is interesting to note
that 
increasing the number of 
grid points (for the same grid spacing) beyond a certain threshold 
leads to the occurrence of spurious eigenvalues.
As shown in Table~\ref{Tab1}, when
increasing the number of grid points for $\tau=0.6$ from $2n=20$ to $30$,
then the size of the Langevin system (\ref{Langevin}) increases 
only by two (and not by five, as one would expect).
The situation is even more striking for $\tau=0.4$: increasing the number of 
data points from $2n=30$ to $44$ leads to only one additional auxiliary 
variable in (\ref{Langevin}). 

\subsection{Molecular dynamics data}
\label{Subsec:MDdata}
In earlier work \cite{JHS17} we have employed an MD
simulation to generate VACF data of a single colloid in    
a Lennard-Jones (LJ) fluid. In that work we have developed a new iterative 
procedure (which we called IMRV method) to determine a memory kernel for a 
generalized Langevin equation of a coarse-grained model.
Here we demonstrate the consistency of the memory kernel (\ref{memory})
corresponding to the extended Langevin model (\ref{Langevin}) based on 
our approximation of the same VACF with the results from \cite{JHS17}.

\begin{figure}
\centerline{\includegraphics[height=5cm]{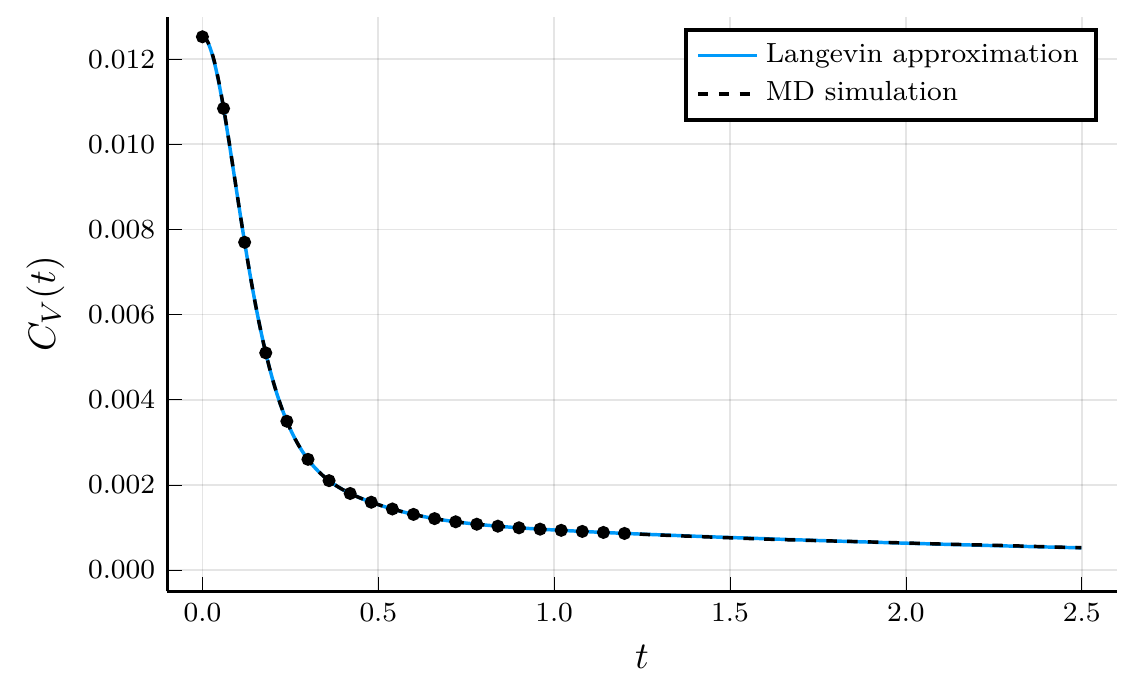}}
\caption{\label{Fig4}VACF from an MD simulation, with (unnormalized)
inter\-polation, points and the corresponding approximation.}
\end{figure}

In preliminary experiments we found that in this setup
the interpolation grid should extend
up to a final time somewhat beyond \mbox{$t=1$}, and therefore we have chosen 
the parameters \mbox{$\tau=0.06$} and \mbox{$n=10$} for this example
(see Figure~\ref{Fig4}). With these parameters we haven't encountered any 
spurious eigenvalues nor negative real ones, so the extended system is also 
of size $10\times 10$ with $N=9$ auxiliary variables. 

\begin{figure}
\centerline{\includegraphics[height=5cm]{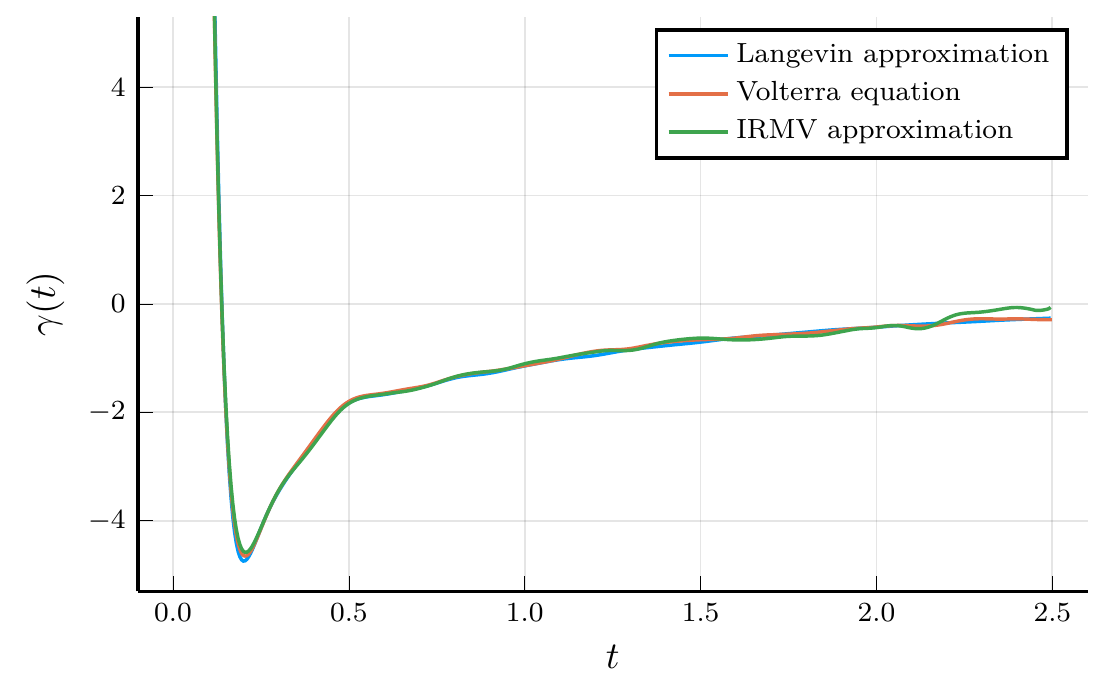}}
\caption{\label{Fig5}Approximations of the memory kernel for the MD simulation.}
\end{figure}

From Figure~\ref{Fig5} it can be seen that the 
corresponding memory kernel approximation (\ref{memory}) is very similar to 
the two kernels computed in \cite{JHS17} with the IMRV method and via the
(second order) Volterra integral equation scheme suggested in \cite{SKTL10}.

\subsection{The impact of data noise}
\label{Subsec:noise}
To assess how well our method performs with noisy data, we 
generated three data sets with shorter run times, which results in 
less smooth VACFs.

The setup of these MD simulations is almost identical to that in the 
previous section. The system is that of a colloid  in a fluid of
$N=31627$ LJ particles with size $\sigma$ and mass $m$. We use truncated
and shifted LJ potentials with the energy scale $\epsilon$ which are cut
off  at $r_{\mathrm{c}}=2^{\frac{1}{6}}\sigma$, resulting in purely
repulsive WCA interactions. The cubic simulation box has periodic 
boundary  conditions in all three dimensions and a side length
$35.76\sigma$. The length, energy, and mass scales in the system are 
defined by the LJ diameter $\sigma$, energy $\epsilon$, and
mass $m$, respectively, which defines the LJ time scale 
$t_{\mathrm{LJ}} = \sigma \sqrt{m/\epsilon}$. The colloid has a mass of $M=80m$ 
and is defined as a rigid body with a radius $R=3\sigma$. In these
simulations the body of the colloid is constructed so that its
surface is more smooth than that of the simulations 
in \cite{JHS17}, resulting in full slip boundary conditions
for the LJ fluid. In order to sample at our desired temperature,
$\beta = 1$, we equilibrate the system with a Langevin 
thermostat. All simulations are performed using LAMMPS \cite{LAMMPS}.

We have generated three data sets with different 
levels of noise corresponding to the following 
production run lengths, where we use a time step of 
$\Delta t_{\mathrm{MD}}=0.001$:
\begin{enumerate}
\item Low noise setting: \\
      $5$\,M steps for the production run;
\item Medium noise setting: \\
      $1$\,M steps for the production run;
\item High noise setting: \\
      $0.1$\,M steps for the production run.
\end{enumerate}

\begin{figure*}
\centerline{\includegraphics[height=5cm]{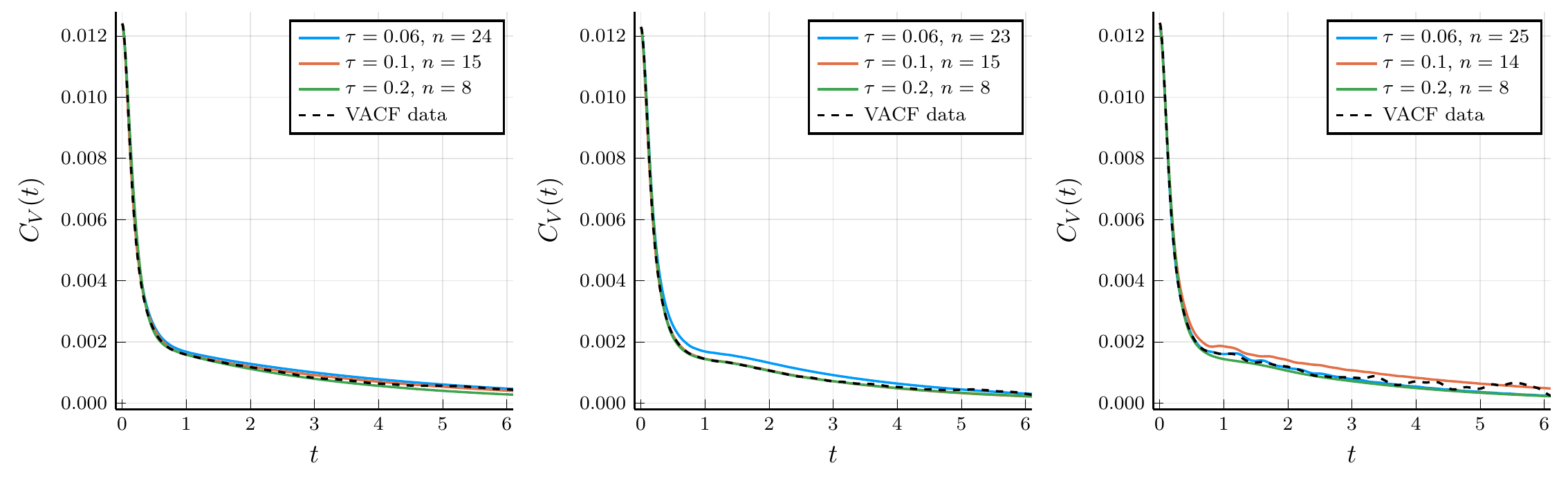}}
\caption{\label{Fig6}VACF approximations for noisy data sets: 
(i) left panel, (ii) middle panel, and (iii) right panel.}
\end{figure*}

In Figure~\ref{Fig6}, it can be seen that the relevant details of the VACF
occur in the time interval $[0,3]$, while data noise
seems to dominate beyond $t=5$ -- in each of the three settings. 
We therefore found it appropriate to sample interpolation data from about the 
time interval $[0,3]$, using three different grid spacings:
\begin{itemize}
\item $\tau=0.06$ with $n=25$ (50 samples),
\item $\tau=0.1$ with $n=15$ (30 samples),
\item $\tau=0.2$ with $n=8$ (16 samples).
\end{itemize}
However, not all transfer functions associated with these data have
been positive real; when they weren't we reduced the dimension $n$ by one,
and retried, until being successful eventually. 

\begin{table}[h]
\footnotesize
\centerline{\begin{tabular}{@{}l|rr|rr|rr}
\br
& \multicolumn{2}{c|}{(i)} & \multicolumn{2}{c|}{(ii)} & 
  \multicolumn{2}{c}{(iii)}\\[1ex]
& $n$ & $N$ & $n$ & $N$ & $n$ & $N$ \\
\br
$\tau=0.06$ & 24 & 12 & 23 & 12 & 25 & 16\\
$\tau=0.1$  & 15 &  5 & 15 &  9 & 14 &  7\\
$\tau=0.2$  &  8 &  3 &  8 &  4 &  8 &  4\\
\br
\end{tabular}}
\caption{\label{Tab2}Grid parameter $n$ and number $N$ of auxiliary variables 
$Z$ in (\ref{Langevin}) for noisy data.}
\end{table}

This is documented in Table~\ref{Tab2}: For example, the entry $n=23$ for the 
medium noise setting (ii) and grid spacing $\tau=0.06$ bears witness 
that both for $n=24$ and $n=25$ the corresponding transfer functions failed 
to be positive real. Overall, out of the $9=3\cdot 3$ cases that we have used
data samples for, six have been successful right away, while three of them
failed initially, namely setting (i) with $\tau=0.06$, setting (ii) with 
$\tau=0.06$, and setting (iii) with $\tau=0.1$. 
It is difficult, though, to deduce a general rule of 
thumb for when the algorithm is prone to fail. Our experience indicates that
the probability increases significantly beyond $n=20$, but the algorithm may
break down for $n=25$, and yet be successful for $n=26$.

\begin{figure*}
\centerline{\includegraphics[height=5cm]{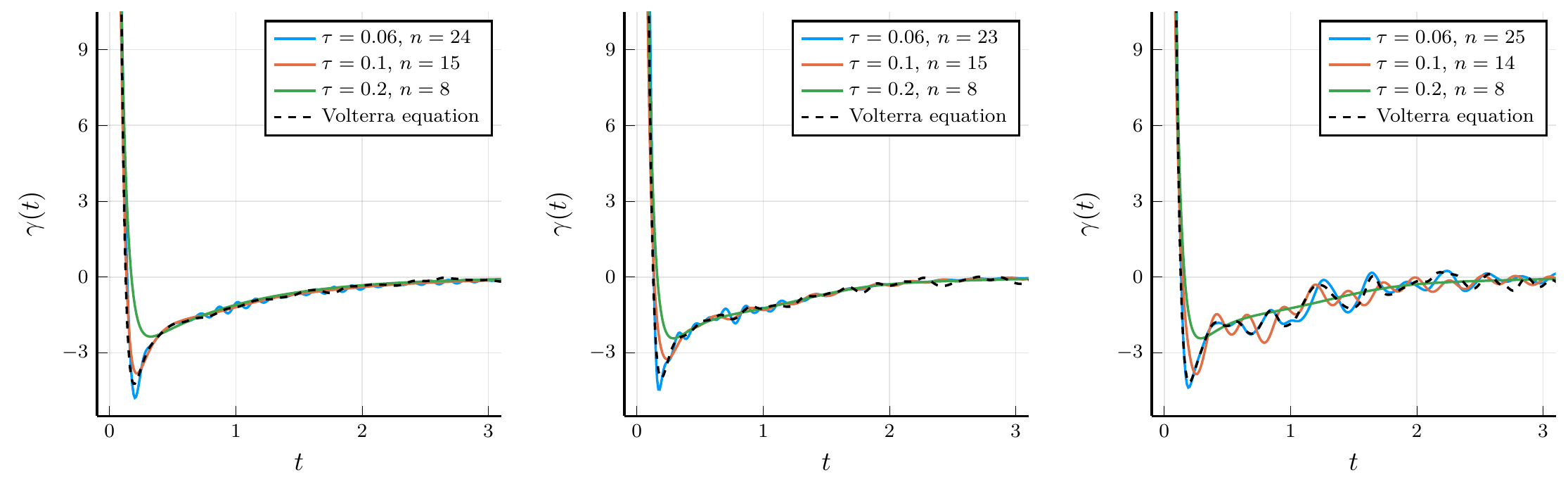}}
\caption{\label{Fig7}memory approximations for noisy data sets: 
(i) left panel, (ii) middle panel, and (iii) right panel.}
\end{figure*}

Turning to the VACF plots in Figure~\ref{Fig6} it can be seen that 
in the low and medium noise settings, (i) and (ii), the VACF approximation
for the grid spacing \mbox{$\tau=0.1$} is best, whereas $\tau=0.2$ yields the
best result in the high noise setting (iii). 
The latter is surprisingly smooth in spite of the high amount of noise,
and looks very similar to the one from setting (ii). 
In fact, according to Table~\ref{Tab2} both approximations are based on only 
five ($=N+1$) non-spurious eigenvalues of $A$ and their associated 
eigenvectors, and those probably carry much the same information in both
settings. 

Overall we conclude that \mbox{$\tau=0.06$} is too small for this MD setup: 
In the high noise setting (iii) the corresponding VACF reconstruction picks up
too much noise and is oscillating a lot; in the other two settings 
the noise doesn't manifest itself in the reconstruction but in a series of
spurious eigenvalues \mbox{($N\ll n-1$)}, the elimination of which
causes a certain loss of the interpolation property and restricts the quality 
of the approximation. This is most striking for the medium noise 
setting (ii) -- which is the example where dimensions $n=25$ and $n=24$ failed.

Concerning the impact of noise on the associated memory terms,
Figure~\ref{Fig7} reveals
that with increasing noise more and stronger oscillations appear in the 
memory kernels. This effect becomes stronger with decreasing
grid width $\tau$, but all Langevin models are more robust in that respect
than the Volterra equation method, which we have employed for ``calibration''.
Taking the kernel calculated with the Volterra method for the small 
noise setting (i) as ``ground truth,''
then one can see that the Langevin models for $\tau=0.2$ fail to
match the depth of the sharp minimum of the kernel near $t=0.2$,
whereas the memory kernels for the other two grids provide a reasonable fit 
of this particular feature, regardless of the noise magnitude.
Apparently, the grid spacing $\tau=0.2$ is
too wide for that purpose -- although this could not be perceived from the
VACF reconstructions. On the other hand, in the high noise setting (iii)
the fairly smooth memory for $\tau=0.2$ is the most convincing one.
We conclude that for larger amounts of noise it pays off to use a larger
grid spacing, and that the quality of the associated memory kernel correlates
well with the quality and the smoothness of the VACF approximation. 

\subsection{Computational efficiency}
\label{Subsec:Performance}

The accurate determination of an extended Langevin model allows
systems to be coarse-grained. However, for such a model to be useful
as a coarse-graining tool, it needs to provide a sufficient speedup to
simulation runtimes.  To evaluate the speedup, we simulate a
coarse-grained model of the system discussed 
in Section~\ref{Subsec:noise} using the
extended Langevin model determined from the method outlined in this
paper. We then compare this runtime to that of a fine-grained MD
simulation of the same system, using the same time step 
$\Delta t_{\mathrm{MD}}=0.001$. All simulations are performed in 3-dimensions using LAMMPS
\cite{LAMMPS}. Although the described method is derived in
one-dimension, each component of the velocity in our system is
uncorrelated.  Therefore, we can apply the determined extended Langevin
model to each component of the velocity and maintain the system
properties. For our coarse-grained simulation, we use an
Euler-Maruyama integration scheme.

As expected, simulations using our extended Langevin model were
significantly less computationally expensive than the 
fine-grained MD simulations. The
speedup factor per time step was $160$ for an extended Langevin
model with $9$ auxiliary variables, and $170$ for a Langevin model
with only $5$ variables. Normalizing these numbers by the degree of
coarse-graining (i.e., the number of fine-grained particles per
coarse-grained particle), these numbers are comparable to those
reported by Wang \emph{et al.}~\cite{WMP20}, who used their machine
learning based auxiliary variable schemes with comparably high number of
auxiliary variables to study dilute star polymer solutions.
We note that the actual speedup per time {\em unit} is even larger due 
to the fact that the time steps in coarse-grained simulations can be chosen 
much larger than in microscopic simulations \cite{JHS17,WMP20}.

\section{Conclusion}
\label{Sec:Conclusion}

In the present work, we have addressed the problem of mapping
GLEs onto equivalent extended Markovian Langevin equations which can be integrated
more easily in numerical simulations. We have developed an analytical
method to extract the parameters of the extended Markovian system
directly from the knowledge of the autocorrelation function of the
target quantity, without knowledge of the memory kernel in the GLE.
The memory kernel can also be evaluated in retrospect 
{\em via} Eq.\ (\ref{memory}). 
Importantly, in contrast to related recent approaches 
based on auxiliary variables \cite{CBP10,BB13,SLK14,MLL16a,WMP20}, 
the number of independent degrees of freedom in our extended 
Markovian  system is the same as that in the original GLE. Thus
we are not extending the configurational space of the system.
Compared to previous methods to derive extended Markovian
integrators for GLEs, our method has the advantage that the 
number of auxiliary variables can be chosen much larger, 
even beyond twenty in some of our test runs.
Therefore, the memory kernel can be represented very 
accurately. Moreover, as we have shown above, the algorithm can 
handle input data that suffer from large statistical noise.

We have derived the algorithm using the example of a GLE for
Brownian dynamics with memory. In doing so, we have assumed that the
derivative of the VACF vanishes at time
zero, $\dot{C}_V(0)=0$. This is certainly the case if it is derived
from an underlying atomistic model with reversible, Hamiltonian
dynamics. However, it may not be correct if the underlying
fine-grained dynamics already has a dissipative component. From a
physical point of view, this would imply that the dynamics of the
system is governed by processes on vastly different time scales --
intermediate time scales that are captured by the memory kernel, and
very short scales that are captured by additional instantaneous 
friction terms. Our algorithm can be extended to account for such a
possibility, as we have demonstrated in Section~\ref{Subsec:slowdiffusion}.

In the current form, the algorithm can be used to integrate
one dimensional GLEs or higher dimensional GLEs, if 
the memory and stochastic part of the high dimensional GLE can be 
written as a sum of independent one-dimensional GLEs. This
is the case in most current applications of Non-Markovian modelling, 
e.g., in GLEs based on self-memory only \cite{WMP20} and in 
Non-Markovian dissipative particle dynamics \cite{LLDK17}.
In the present form, this method cannot yet be
applied to general multidimensional GLEs with pair memory \cite{JHS19}.
Extending the algorithm for such cases will be the subject 
of future work.

\ack

We thank Viktor Klippenstein, Madhusmita Tripathy, and Nico van der
Vegt for fruitful discussions.  The research leading to this work has
been done within the Collaborative Research Center SFB TRR 146;
corresponding financial support was granted by the Deutsche
Forschungsgemeinschaft (DFG) via Grant 233530050.  GJ also gratefully
acknowledges funding by the Austrian Science Fund (FWF): I 2887.
Computations were carried out on the Mogon Computing Cluster at ZDV
Mainz.

\appendix
\section{The Newton scheme}
\label{App:Newton}
In order to satisfy (\ref{A11}), we consider the (1,1)-entry $A_{1,1}$
of $A$ as a function $\varphi$ of $y_1$, choose $y_1^{(0)}=y_1$ as initial guess,
and use the Newton iteration
\begin{equation}
\label{Newton}
   \ y_1^{(m+1)} \!= 
   y_1^{(m)} - \frac{\varphi(y_1^{(m)})}{\varphi'(y_1^{(m)})}\,,
   \ \, \nu=0,1,\dots,
\end{equation}
until a suitable value $y_1^{(m+1)}$ with
$\varphi(y_1^{(m+1)})\approx 0$ has been found. 
However, instead of the target function $\varphi(y_1)=A_{1,1}$, 
chosen in (\ref{A11}) for the ease of presentation, we (need to) use
\begin{equation}
\label{phi}
   \ \varphi(y_1) \,=\, {\rm Re}\,A_{1,1}
\end{equation}
in (\ref{Newton}).
The reason is that negative eigen\-values of $J$ result in a nonzero 
imaginary part of $A$, and it is only the real part which is relevant for our 
purpose because of the way we subsequently modify these eigenvalues, 
see \ref{App:SpectralModifications}.

The difficulty in (\ref{Newton}) is the evaluation of the derivative 
$\varphi'$ because of the nontrivial recursion that is used to set
up the tridiagonal matrix $J$. We resolve this problem in two steps. 
First, we use algorithmic differentiation~\cite{GW08} to simultaneously 
determine $J$ and its derivative
\begin{equation}
\label{E}
  \ \partial J \,=\, \frac{\rmd J}{\rmd y_1} \, \in \R^{n\times n}.
\end{equation}
Second, we employ the chain rule
\[
  \ \partial A \,:=\, \frac{\rmd A}{\rmd y_1}
  \,=\, \frac{1}{\tau}\,
        \frac{\rmd}{\rmd J}\!\log(J)\,\, \partial J
\]
and the useful relation
\begin{equation}
\label{Higham}
  \ \log\Bigl(\begin{cmatrix} J & \partial J \\ 0 & J \end{cmatrix}\Bigr)
  \,=\, \begin{cmatrix} A & \partial A \\ 0 & A\phantom{'} \end{cmatrix},
\end{equation}
which follows from
\cite[p.~58]{Higham08}.

This yields the expression
\[
  \ \varphi'(y_1) \,=\, {\rm Re}\,(e_1^T\partial A\,e_1) 
    \,=\, {\rm Re}\,\partial A_{1,1}\,,
\]
to be used in (\ref{Newton}), see \ref{App:Algorithm} for further details.

We finally mention that there is no need to determine $y_1$ to high accuracy, 
because (i) the subsequent modifications of $A$,
cf.~\ref{App:SpectralModifications}, will slightly perturb this matrix 
entry anyway, and (ii) because of the way we (approximately) solve the 
singular Lur'e equations, cf.~\ref{App:Algorithm}. 
In our experiments two to seven Newton steps were always sufficient.

\section{Spectral modifications of the system matrix}
\label{App:SpectralModifications}

\subsection*{Eliminating spurious exponentials}
Given the eigenvector matrix (\ref{X}) of $A$, 
we let $z=X^{-1}e_1=[z_1,\dots,z_n]^T$.
Then we readily conclude from (\ref{A}) the
Prony series representation
\begin{equation}
\label{f-App}
  \ f(t) \,=\, e_1^TX e^{t\Lambda} X^{-1}e_1
  \,=\, \sum_j w_j e^{t\lambda_j}
\end{equation}
of (\ref{expinterpolsol}) with
\begin{equation}
\label{omega}
  \ w_j = x_{1j}z_j\,,
\end{equation}
where $x_{1j}$ is the first entry of the $j$th eigenvector $x_j$.

Let us now make the assumption that $\lambda_n$ is a spurious eigenvalue of $A$
in the closed right-half plane, which we want to eliminate from (\ref{f-App}),
because it is non-physical and has negligible impact on $f$
for $t\in\triangle$, since $|x_{1n}|$ and $|z_n|$ are sufficiently small. 

We denote by $\Lambda'$ the diagonal matrix with the eigenvalues
$\lambda_1,\dots,\lambda_{n-1}$ of $A$. Then we delete the final column of $X$ 
and select one out of rows $2$ to $n$, which we also delete from $X$ 
to obtain an invertible matrix $X'\in\R^{(n-1)\times(n-1)}$ --
in our code we always took the last row. 
Finally, let $z'=[z_1,\dots,z_{n-1}]^T$
and $e_1'$ denote the first canonical basis vector in $\R^{n-1}$. 
Since $z_n$ is assumed to be negligible, it follows that
\[
  \ X'z' \,\approx\, e_1'\,, \quad {\rm i.e.}, \quad 
  z'\,\approx\,{X'}^{-1}e_1'\,,
\]
and hence, for $t\in\triangle$ there holds
\[
  \ {e_1'}^{\!T}X'e^{t\Lambda'}{X'}^{-1}e_1' 
  \,\approx\, {e_1'}^{\!T}X'e^{t\Lambda'}z'
  \,=\, \sum_{j=1}^{n-1}w_j e^{t\lambda_j}
\]
by virtue of (\ref{omega}). We therefore achieve our goal by replacing $A$ by 
\begin{equation}
\label{Aprime}
  \ A' \,=\, X'\Lambda'{X'}^{-1}\,.
\end{equation}

\subsection*{Treatment of negative eigenvalues of $J$}
Recall that we have made the assumption that all eigenvalues of $J$ are
different. Let us now assume that the eigenvalue $\mu_n$ of $J$ 
belongs to the interval $(-1,0)$; it comes with a real eigenvector $x_n$
and a real weight $w_n$ in (\ref{f-App}). 

In order to replace the corresponding exponential term by the one 
in (\ref{negative-term}), we define the diagonal matrix
\[
  \ \Lambda' \,=\,
{\arraycolsep0.3ex
    \begin{cmatrix} \lambda_1 & \\[-1ex] & \ddots & \\[-1ex]
                    & & \lambda_n \\[-0.5ex]  & & & \lambda_{-n} \end{cmatrix}
}
\]
with $\lambda_{\pm n}$ as in (\ref{lambda-special}), and we further let
\[
  \ X' \,=\,
    \begin{cmatrix} x_1 & \cdots & x_{n-1} & x_n & x_n \\
                    0 & \cdots & 0 & \rmi & -\rmi
    \end{cmatrix}
\]
and
\[
  \ z' \,=\, 
  \Bigl[ z_1 \ \cdots \ z_{n-1} \ \frac{z_n}{2} \ \frac{z_n}{2} \Bigr]^T.
\]
The matrix $X'$ is an invertible $(n+1)\times(n+1)$ matrix, and it satisfies
\[
  \ X'z' \,=\, \begin{cmatrix} Xz \\ 0 \end{cmatrix} 
  \,=\, \begin{cmatrix} e_1 \\ 0 \end{cmatrix} \,=\, e_1'\,,
\]
where $e_1'$ is the first canonical basis vector in $\R^{n+1}$. 
It therefore follows that
\begin{eqnarray*}
  \ {e_1'}^{\!T}X'e^{t\Lambda'}{X'}^{-1}e_1' 
  \,=\, {e_1'}^{\!T}X'e^{t\Lambda'}z' \\
  \ \,=\, \sum_{j=1}^{n-1}w_j e^{t\lambda_j}
          \,+\, \frac{1}{2}\,w_n e^{t\lambda_n}
          \,+\, \frac{1}{2}\,w_n e^{t\lambda_{-n}}\\
  \ \,=\, \sum_{j=1}^{n-1}w_j e^{t\lambda_j} 
          \,+\, w_n|\mu_n|^{t/\tau}\!\cos(\pi t/\tau)
\end{eqnarray*}
by virtue of (\ref{lambda-special}).
Since this is the Prony series we have been looking for, we achieve our goal 
by replacing $A$ by
\begin{equation}
\label{Aprime2}
  \ A' \,=\, X'\Lambda'{X'}^{-1}\,.
\end{equation}

\section{Outline of the full algorithm}
\label{App:Algorithm}

\begin{figure*}[htb]
\centering
\begin{minipage}{.9\linewidth}
\begin{algorithm}[H]
\caption{\ Input: $y_\nu = C_V(\nu\tau)/C_V(0)$, $\nu=0, \ldots, 2n-1$}
\label{Alg:Prony}
\footnotesize
\begin{algorithmic}[1]
\medskip
\State call Algorithm~\ref{Alg:Lanczos} to compute the matrix $J$ and 
       its derivative $\partial J$ with respect to $y_1$
\For{$i = 1, \ldots, i_{\max}$}
\medskip
   \State set $\begin{cmatrix} A & \partial A \\ 0 & A \end{cmatrix} 
               \leftarrow 
               {\displaystyle\frac1\tau}
               \log\begin{cmatrix} J & \partial J \\ 0 & J\end{cmatrix}$
\medskip
   \State set $y_1 \leftarrow y_1 - {\rm Re}\,A_{1,1}/{\rm Re}\,\partial A_{1,1}$
   \State call Algorithm~\ref{Alg:Lanczos} to recompute $J$ and $\partial J$
with the new data vector $y$
\EndFor
\If{$|A_{1, 1}| > \varepsilon$}
   \State error (Newton's method did not find a zero of the function 
          $y_1 \mapsto A_{1, 1}$)
\EndIf
\bigskip
\State factorize $J = XDX^{-1}$ with a diagonal matrix $D$; 
       $[\mu_1,\ldots, \mu_n] \leftarrow {\rm diag}\,(D)$
\State duplicate eigenvalues $\mu_j \in (-1, 0)$ and remove eigenvalues 
       $\mu_j$ with $|\mu_j| \geq 1$; \newline 
       adjust $X$ and $D$ accordingly (see \ref{App:SpectralModifications})
\State set $\Lambda \leftarrow \frac{1}{\tau}\log(D)$; for duplicated 
eigenvalues $\mu_j\in(-1,0)$ employ formula (\ref{lambda-special})
\State set $A \leftarrow X\Lambda X^{-1}$ 
\bigskip
\State set $A_{1,1} = -\delta$ with $\delta > 0$ small (e.g., $\delta=10^{-5}$),
so that $A = \begin{cmatrix} -\delta & b^T \\ -c & A_0 \end{cmatrix}$
\State solve the Riccati equation (\ref{Riccati})
for the symmetric and positive semidefinite matrix $\Sigma_0$
\medskip
\State set $\Sigma \leftarrow 
            \begin{cmatrix} 1 & 0 \\ 0 & \Sigma_0 \end{cmatrix}$, \quad
           $L \leftarrow 
            {\displaystyle \frac{1}{\sqrt{2\delta\beta m}}}
            \begin{cmatrix} 2\delta \\ c-\Sigma_0 b \end{cmatrix}$
\medskip
\If{$A\Sigma+\Sigma A^T \neq - \beta m\,LL^T$}
   \State error (transfer function is not positive real)
\EndIf
\bigskip
\State factorize $A = UTU^{-1}$ (Lanczos algorithm) with 
\[ 
       \ T = \begin{cmatrix} -\delta & ke_1^T \\ -ke_1 & T_0' \end{cmatrix}
       \quad \mbox{tridiagonal and} \quad
       U = \begin{cmatrix} 1 & 0 \\ 0 & U_0 \end{cmatrix}
\]
\bigskip
\State \Return $T$ and $G=U^{-1}L$
\end{algorithmic}
\end{algorithm}
\end{minipage}
\end{figure*}

\begin{figure*}[htb]
\centering
\begin{minipage}{.9\linewidth}
\begin{algorithm}[H]
\caption{\ Input: $y_\nu$, $\nu=0, \ldots, 2n-1$; \quad
         $\Phi$ and $\partial\Phi$ are defined in (\ref{Phi}) and (\ref{dPhi})} 
\label{Alg:Lanczos}
\footnotesize
\begin{algorithmic}[1]
\medskip
\State set $u_{-1} \leftarrow 0$\,, \quad $\partial u_{-1} \leftarrow 0$
\State set $u_0 \leftarrow 1/\sqrt{|y_0|}$\,, \quad
       $\partial u_0 \leftarrow 0$
\State set $\alpha_0 \leftarrow y_1/y_0$\,, \quad 
       $\partial \alpha_0 \leftarrow 1/y_0$
\State set $\gamma_0 \leftarrow 0$\,, \quad $\partial \gamma_0 \leftarrow 0$
\State set $\sigma_0=1$
\medskip
\State set $J_{1, 1} \leftarrow \alpha_0$\,, \quad 
       $\partial J_{1, 1} \leftarrow \partial\alpha_0$
\bigskip
\For{$i = 1, \ldots, n-1$}
   \State set $\tilde u_i \leftarrow xu_{i-1} - \alpha_{i-1}u_{i-1} 
               - \sigma_{i-1}\gamma_{i-1}u_{i-2}$ 
   \State set $\partial \tilde u_i \leftarrow x\partial u_{i-1} 
        - \bigl(\partial\alpha_{i-1} u_{i-1} + \alpha_{i-1} \partial u_{i-1}\bigr)
        - \sigma_{i-1} 
          \bigl(\partial\gamma_{i-1} u_{i-2} + \gamma_{i-1} \partial u_{i-2}\bigr)$
   \vspace{0.3ex}
   \State set $\gamma_i \leftarrow \sqrt{|\Phi[\tilde u_i^2]|}$
   \If{$\gamma_i = 0$}
      \State error (Lanczos algorithm break down)
   \EndIf
   \State set $\partial \gamma_i \leftarrow 
          {\displaystyle \frac{{\rm sign}(\Phi[\tilde u_i^2])}{2\gamma_i}}
           \bigl(
              \Phi[2\tilde u_i\partial\tilde u_i] + \partial\Phi[\tilde u_i^2]
           \bigr)$
   \State set $u_i \leftarrow \tilde u_i/\gamma_i$\,, \quad 
              $\partial u_i \leftarrow  
               {\displaystyle
                \frac{\partial\tilde u_i \gamma_i - \tilde u_i\partial\gamma_i}
                     {\gamma_i^2}}$
   \State set $\alpha_i \leftarrow \Phi[xu_i^2]/\Phi[u_i^2]$
   \vspace{1ex}
   \State set
      $\partial \alpha_i \leftarrow 
       {\displaystyle
        \frac{\Phi[u_i^2]\Phi[2xu_i\partial u_i] 
              + \Phi[u_i^2]\partial\Phi[xu_i^2]
              - \Phi[xu_i^2] \Phi[2u_i\partial u_i]
              - \Phi[xu_i^2] \partial\Phi[u_i^2]}{\Phi[u_i^2]^2}}$
   \State set $\sigma_i \leftarrow \Phi[u_i^2]/\Phi[u_{i-1}^2]$
   \medskip
   \State set $J_{i+1, i+1} \leftarrow \alpha_{i}$\,, \quad
       $\partial J_{i+1, i+1} \leftarrow \partial\alpha_{i}$
   \State set $J_{i+1, i} \leftarrow \sigma_{i}\gamma_{i}$\,, \quad 
       $\partial J_{i+1, i} \leftarrow \sigma_{i}\partial\gamma_{i}$
   \State set $J_{i, i+1} \leftarrow \gamma_{i}$\,, \quad 
       $\partial J_{i, i+1} \leftarrow \partial\gamma_{i}$
\EndFor
\bigskip
\State \Return $J$, $\partial J$
\end{algorithmic}
\end{algorithm}
\end{minipage}
\end{figure*}

We provide a pseudocode formulation of the overall algorithm in 
Algorithm~\ref{Alg:Prony}. For the ease of presentation it comes with a small 
threshold parameter $\varepsilon$ and a maximum number $i_{\max}$ of iterative 
steps to control the Newton iteration. 

In contrast to our presentation in Section~\ref{Subsec:PRL} we have taken 
the liberty to introduce in line 14 of Algorithm~\ref{Alg:Prony}
a negative entry $A_{1,1}=-\delta$ of small absolute value
to circumvent the singular Lur'e equations (\ref{Lure}); this is a kind of regularization.
The associated Lyapunov equation $A\Sigma+\Sigma A^T=-\beta m\,LL^T$
with $L$ as in line 16 is equivalent to the more amenable 
(regular) Lur'e equations
\begin{equation}
\label{Lure2}
\eqalign{\ A_0\Sigma_0+\Sigma_0A_0^T \,=\, -\beta m\,\ell\ell^T\,,\cr
         \ c-\Sigma_0b \,=\, \sqrt{2\delta\beta m}\,\ell\,,}
\end{equation}
for $\Sigma_0$ and $\ell$, cf.~\cite{Anderson67b,AV73}. 
It can be shown that for $\delta\to 0$ the solution of
(\ref{Lure2}) converges to the solution of the corresponding singular
Lur'e system~(\ref{Lure}). Note that when using the second equation
in (\ref{Lure2}) to eliminate $\ell$ from the first one, then this yields
the quadratic Riccati equation 
\begin{equation}
\label{Riccati}
   \ B\Sigma_0 + \Sigma_0B^T + \Sigma_0bb^T\Sigma_0 + cc^T = 0\,, \quad
   B = 2\delta A_0 - cb^T\,,
\end{equation}
for $\Sigma_0$, which is to be solved in line 15.
We emphasize that there exists standard software for the solution of this
equation, and the same is true for the Lanczos algorithm employed in 
line 20.

One consequence of introducing the regularization parameter $\delta$ 
is that the system (\ref{Langevin}) changes into
\[
   \ {\rm d}\!\begin{cmatrix} \Vtilde \\ Z \end{cmatrix}
   \,=\, T \begin{cmatrix} \Vtilde \\ Z \end{cmatrix}\dt
       + G\,\dW\,,
\]
where both $T_{1,1}$ and $G_1$ are nonzero, but small in absolute value.

Before going on let us define, for polynomials
\[ 
   \ p(x) \,=\, \sum_{\nu=0}^{2n-1} a_\nu x^\nu\,, \quad a_\nu\in\R\,,
\]
of degree $2n-1$, at most, the moment functional
\begin{eqnarray}
\label{Phi}
   \ \Phi[p] \,=\, \sum_{\nu=0}^{2n-1} a_\nu y_\nu\,,
\end{eqnarray}
with $y_\nu$ the given data~(\ref{yk}). 
Algorithm~\ref{Alg:Lanczos} determines the corresponding 
Lanczos polynomials $u_i$ of degree $i=0,\dots,n-1$, respectively, 
given by
\[
   \ \Phi[u_iu_j] \,=\, \pm \delta_{ij}\,, \quad i,j=0,\dots,n-1\,,
\]
and returns the Jacobi matrix $J$ with the associated recursion coefficients
and its derivative $\partial J$ with respect to $y_1$. 
For the computation of this derivative we also introduce the 
functional $\partial\Phi$ which maps $p$ as above onto
\begin{eqnarray}
\label{dPhi} 
   \ \partial\Phi[p] \,=\, a_1\,.
\end{eqnarray}

Note that all variables in Algorithm~\ref{Alg:Lanczos} with a prefix $\partial$
denote the derivative with respect to $y_1$ of the same variable without prefix.
Further note that $u_i$, $\tilde u_i$, $\partial u_i$ and $\partial\tilde u_i$
are all polynomials of degree $i$ in the real variable $x$, and that, 
for example, 
$xu_i^2$ is short-hand notation for the polynomial $x\mapsto x(u_i(x))^2$.
We finally mention that in line~18 the variable $\sigma_i\in\{\pm 1\}$, 
and hence, its derivative with respect to $y_1$ is zero.



\end{document}